\documentclass[12pt,a4paper]{article}
\usepackage[T1]{fontenc}
\usepackage[utf8]{inputenc}
\usepackage{lmodern}
\usepackage[margin=27mm]{geometry}
\usepackage{amsmath,amssymb,amsthm,mathtools}
\usepackage{booktabs,tabularx,array}
\usepackage{enumitem,microtype,needspace}
\usepackage{tikz}
\usetikzlibrary{arrows.meta}
\usepackage[authoryear,round]{natbib}
\usepackage[hidelinks]{hyperref}
\setlist[enumerate]{itemsep=4pt,topsep=5pt}
\newcolumntype{Y}{>{\raggedright\arraybackslash}X}
\newcolumntype{P}[1]{>{\raggedright\arraybackslash}p{#1}}
\newtheorem{theorem}{Theorem}
\newtheorem{proposition}{Proposition}
\DeclareMathOperator{\E}{\mathbb E}
\newcommand{\PP}{\mathbb P}
\newcommand{\ind}{\mathbf 1}
\newcommand{\pL}{p_L}
\newcommand{\pH}{p_H}
\newcommand{\vH}{v_H}
\newcommand{\AH}{A_H}
\newcommand{\IH}{I_H}
\hypersetup{pdftitle={Reputation and Delegated Public Spending},pdfauthor={Emin Ablyatifov and Georgy Lukyanov}}
\title{Reputation and Delegated Public Spending\thanks{We thank Arthur Izgarshev, David Li, Anastasia Makhmudova, Konstantin Shamruk, Yuliia Tukmakova, Anna Vlasova, and Stanislav Vyushkov for valuable comments. All remaining errors are our own.}}
\author{Emin Ablyatifov\thanks{Geneva Business School, Geneva, Switzerland.
Email: \href{mailto:eablyatifov@gbsge.com}{eablyatifov@gbsge.com}.}
\and Georgy Lukyanov\thanks{Toulouse School of Economics, Toulouse, France.
Corresponding author. Email:
\href{mailto:georgy.lukyanov@tse-fr.eu}{georgy.lukyanov@tse-fr.eu}.}}
\date{27 October 2026}

\begin{document}
\maketitle

\begin{abstract}
This paper studies a repeated funding relationship between a grant administrator and an implementing agency. The agency is either committed to implementation or able to divert the advance. The administrator chooses whether to fund and whether to verify, but cannot commit to future decisions. We construct a stationary equilibrium in which agencies with intermediate reputations are audited. An inspection either reveals diversion and ends funding or leads to continued funding with intermittent auditing. In the latter case, audits sustain implementation without further revealing type, while a strategic agency diverts unaudited advances. Cheaper auditing makes funding worthwhile at lower reputations. For initial beliefs audited under both cost levels, it raises the administrator's value but reduces expected long-run implemented spending: a higher reputation is required to retain funding, and fewer strategic agencies meet that requirement. For beliefs audited under both programs, better detection at unchanged cost instead raises implementation without changing the probability of continued funding. These results show why project delivery alone need not measure the value of improved monitoring.
\end{abstract}



\section{Introduction}\label{sec:intro}

A grant administrator that advances funds to an implementing agency
faces two related decisions: whether to finance the next project and
how closely to monitor the use of the advance. Monitoring affects the
agency's current incentives and the information on which subsequent
funding depends. A change in monitoring technology can therefore alter
both the behavior of funded agencies and the composition of those that
continue to receive funds.

Repeated advances provide a setting in which these effects matter.
UNDP's procedures require audit and spot-check findings to be considered
when deciding whether to withhold or reduce subsequent advances
\citep{UNDP}. The HACT framework similarly connects assurance activities
and transfer arrangements to the risk associated with an implementing
partner \citep{HACT}. Monitoring also affects the use of current funds:
\citet{Olken2007} finds that an announced increase in audit probability
reduced missing expenditures in village road projects, measured against
independent engineering estimates. These observations motivate our
focus on repeated funding and anticipated verification. They do not
establish the particular information structure or reputational
mechanism studied here.

This paper develops a model in which an administrator repeatedly funds
the same agency. The agency is either committed to implementation or
strategic, in which case it can divert the advance. In each period, the
administrator can withhold funding, fund without verification, or use a
specified verification program. The agency observes this choice before
acting. An inspection reveals the action exactly; otherwise, no
information arrives. Both players are long-lived, and future funding
and monitoring are discretionary. The advance and the available program
are fixed, and there are no contingent transfers.

Our main result is that cheaper auditing can increase the
administrator's value while reducing long-run implemented spending.
In the equilibrium we construct, a fall in audit cost makes it
worthwhile to fund agencies with lower initial reputations. At the
same time, it raises the reputation required for continued funding with
intermittent auditing. Among agencies whose initial reputations are
audited under both cost levels, fewer strategic agencies satisfy this
requirement and retain funding. Every committed agency retains funding,
so implementation per funded agency rises even as total implementation
in the initial population falls.

The incentive to implement comes from future access to advances. At
intermediate reputations the administrator always uses the verification
program, and a strategic agency randomizes between implementation and
diversion. The first inspection either reveals diversion and ends
funding or verifies implementation and raises reputation to the
threshold at which the administrator is willing to relax monitoring.
At this threshold, the administrator randomizes between audited and
unaudited funding. The strategic agency implements under the program
and diverts otherwise. The prospect of future unaudited advances
compensates it for foregoing diversion when the program is in place.

Audits continue after this threshold is reached, but they no longer
reveal type. Both types implement under the program, and unaudited
funding produces no report. A strategic agency can consequently receive
an indefinitely long sequence of clean findings while diverting
unaudited advances. The administrator anticipates this behavior: a
clean finding verifies the inspected project without providing further
evidence that the agency is committed. We characterize the resulting
limiting distribution of reputation and long-run implementation.

A lower audit cost and a higher detection probability have different
effects. Better detection at unchanged expected cost raises long-run
implementation without changing the probability of continued funding
for agencies initially audited under both policies. It brings the first
informative inspection forward and reduces unaudited funding thereafter.
If additional inspections must be paid for, however, greater detection
can lower the administrator's value. These comparisons are between
available verification programs, each with a specified cost and
detection probability.

The paper relates to work on oversight and relational incentives.
\citet{AK2025} study resource approval and costly oversight when effort
changes persistent quality, and also show that cheaper oversight can
weaken incentives. In their model, more frequent inspections shorten
periods of trust and can eliminate the agent's incentive to invest.
In our equilibrium, cheaper verification leaves the strategic agency's
continuation rent and implementation frequency at the trust threshold
unchanged. It raises the threshold itself, reducing the probability
that a strategic agency reaches it. The comparison therefore concerns
selection among fixed types as well as the distinction between
implemented spending and the administrator's value. \citet{AK2024} show that relational enforcement
can require inspections whose equilibrium outcome is already known.
In our setting, access to future advances sustains implementation, and
funding decisions based on reputation determine which agencies keep
that access.

The model also relates to reputation with endogenous information.
\citet{Liu2011} studies costly learning about past transactions with a
long-lived firm, while \citet{Garcia2011} considers monitoring an
informed agent's reports when the agent's preferences are private and
employment cannot be terminated. Here,
verification concerns current implementation and affects both continued
funding and subsequent project delivery. In the dynamic verification
literature, \citet{MQ2005} characterize optimal contracts specifying
verification and contingent transfers, including an infinite-horizon
extension. Our analysis concerns equilibrium with restricted funding
instruments and limited commitment.\footnote{For a static analysis of
auditing and contingent allocations, see \citet{MP1989}. The dependence
of current behavior on future treatment also underlies the ratchet
effect \citep{Weitzman1980}. Here the incentive comes from future access
to advances; performance targets are not revised.}

Implemented spending measures delivery within this funding
relationship. It differs from the institutional fiscal capacity studied
by \citet{BP2009}, which involves taxation and investment in state
capabilities. We take the administrator's ability to raise and transfer
funds as given and study their use by the implementing agency.

Section~\ref{sec:model} presents the model, and
Section~\ref{sec:equilibrium} constructs the equilibrium.
Section~\ref{sec:longrun} characterizes learning and implementation.
Section~\ref{sec:policy} compares changes in audit cost and detection.
Section~\ref{sec:discussion} discusses the assumptions on monitoring
and information and concludes. The appendix contains the longer proofs
and extensions.

\section{The model}\label{sec:model}

\subsection{Projects, preferences, and timing}

Time is discrete and infinite, $t=0,1,\ldots$. A grant administrator
faces the same implementing agency in every period. The agency privately
knows its fixed type. Type $H$ is committed to implementing every funded
project. Type $O$ chooses between implementation and diversion. The
common initial probability of type $H$ is $p_0\in(0,1)$. Both players
discount future payoffs by $\beta\in(0,1)$.

A project requires an advance $R>0$. Implementation produces a public
benefit $b>0$ and gives the strategic agency no private payoff. Diversion
produces no public benefit and gives that agency a private payoff $R$.
We interpret diversion as deliberate private appropriation of a
project advance. Funding costs the administrator $k>0$, including the
opportunity cost of committing the funds. The administrator values project benefits net of funding and
monitoring costs and places no weight on diversion receipts. We use
its discounted payoff to assess the value of a verification program.
This objective differs from the sum of the players' payoffs, which
would also count the agency's diversion receipts.\footnote{The advance
$R$ and its cost to the administrator $k$ are separate fixed
parameters. The size of the project, the cost of raising funds, and
the private value of diversion are taken as given. A payoff received
in every period independently of funding, actions, and reports can
be normalized away. Compensation that ends when funding stops or
depends on verified performance would change incentives and is outside
the model.}

The administrator has three current choices: no funding, denoted by $0$;
funding without verification, denoted by $N$; and funding under a
specified verification program, denoted by $A$. Program $A$ costs $c>0$
in expectation and inspects the project with probability
$\tau\in(0,1]$. An inspection reveals implementation or diversion
exactly. The timing within a period is as follows.
\begin{enumerate}
\item The administrator chooses $0$, $N$, or $A$. The agency observes
the realized choice. A lottery over choices is resolved at this stage.
\item If funded, type $O$ chooses implementation or diversion. Type
$H$ implements. Diversion pays $R$ before any inspection result.
\item Under $A$, an independent inspection draw is realized after the
action. The public report records implementation, diversion, or no
inspection. Under $0$ and $N$, there is no report.
\item The public belief is updated and the next period begins.
\end{enumerate}

The administrator can commit to the current program, but future
funding and monitoring remain discretionary. When the agency acts, it
knows whether the program is in place; the inspection draw occurs
afterwards. Thus an uninspected project under $A$ carries a risk of
detection at the time of the agency's choice, whereas a project under
$N$ does not.

We interpret $A$ as a prespecified verification package for a fixed
project. The administrator can use this package or fund without it,
but cannot purchase an intermediate detection probability. Grant
systems may also use financial reports, corrective measures, and
alternative transfer arrangements. We restrict the available
instruments to study how future access to advances alone can induce
implementation.

The administrator does not independently observe implementation or
the benefit received by project beneficiaries. In particular, it does
not learn its realized payoff from an unaudited project. The same
agency remains in place throughout: there is no replacement, recovery
of diverted funds, fine, delayed inspection, or independent news about
type. These restrictions make verification the only source of new
information and continued funding the only reward for implementation.

We impose the following sufficient restrictions throughout the baseline:
\begin{equation}
 b>k>0,\qquad
 \beta\tau>1-\beta,\qquad
 0<c<\min\left\{b-k,\frac{\beta\tau k}{1-\beta}\right\}.
 \label{eq:domain}
\end{equation}
The patience condition makes future access to advances valuable enough
to support implementation. The cost bounds allow audited implementation
to generate positive surplus and make audited funding worthwhile at
some beliefs where funding without verification would be rejected.
The restrictions describe a nonempty region: for any $\beta>1/2$,
one can choose $\tau>(1-\beta)/\beta$ and then a sufficiently small
positive $c$. They are sufficient for the equilibrium below; other
equilibria may exist outside this region.

\subsection{Beliefs and equilibrium}

Let $p$ be the public posterior probability of type $H$ before the
administrator's choice. Let $\mu$ denote the conjectured probability
that the strategic agency implements under $A$. Thus
\begin{equation}
 q=p+(1-p)\mu
 \label{eq:q}
\end{equation}
is the probability of implementation under the program. Here $\mu$ is the probability with which the strategic agency chooses
implementation. Conditional on that choice, the project succeeds.

No funding, funding under $N$, and a realization of no inspection under
$A$ leave the posterior unchanged. Under $A$, verified implementation
gives
\begin{equation}
 T_D(p;\mu)=\frac{p}{p+(1-p)\mu},
 \label{eq:bayes}
\end{equation}
when the denominator is positive. Verified diversion gives posterior
zero, since the committed type never diverts. The administrator has no
private information about type, so its current choice conveys no
additional information, including after a deviation from its prescribed
policy.

Two conventions specify beliefs at otherwise unreached observations.
At $p=0$, verified implementation leaves the posterior at zero. Verified
diversion is assigned posterior zero even when the prescribed actions
make that event impossible. The latter convention respects the
restriction that only the strategic type can divert.

We construct a stationary perfect Bayesian equilibrium. The
administrator's policy depends on $p$, and the strategic agency's
response depends on $p$ and the observed current choice. Beliefs follow
Bayes' rule with the conventions above. We specify an optimal agency
response after every feasible choice, including one that the
administrator does not make in equilibrium. In checking the agency's
incentives, we hold the conjecture in \eqref{eq:bayes} fixed: an
unobserved deviation changes the probabilities of reports, while the
administrator continues to interpret each report using the prescribed
strategy. The constructed strategies use fresh independent draws at
each randomization; deviations may depend on the entire history.

\section{Equilibrium funding and verification}\label{sec:equilibrium}

\subsection{Cutoffs and strategies}

Write $\delta=1-\beta$ to shorten the formulas, and define
\begin{align}
 \kappa&=\delta+\beta\tau,&
 \pH&=1-\frac cb,&
 \theta&=\frac{\delta}{\beta\tau},\notag\\
 \vH&=\frac{b-k-c}{\delta},&
 \AH&=b+\beta\tau\vH,&
 \pL&=\frac{\pH(k+c)}{\AH}.
 \label{eq:constants}
\end{align}
The restrictions in \eqref{eq:domain} imply
\begin{equation}
 0<\pL<k/b<\pH<1,\qquad 0<\theta<1.
 \label{eq:ordering}
\end{equation}
The lower cutoff $\pL$ determines whether funding begins. At the
upper cutoff $\pH$, the administrator is willing to relax verification;
we call it the trust threshold. The probability $\theta$ will determine
how often an agency at that threshold receives an unaudited advance.
The remaining constants simplify the value functions below.

The administrator's policy is
\begin{equation}
 \sigma(p)=
 \begin{cases}
 0,&0\leq p\leq\pL,\\
 A,&\pL<p<\pH,\\
 N\text{ with probability }\theta,
 \ A\text{ with probability }1-\theta,&p=\pH,\\
 N,&\pH<p\leq1.
 \end{cases}
 \label{eq:policy}
\end{equation}
The strategic agency diverts under $N$. Its implementation probability
after $A$ is
\begin{equation}
 m(p)=
 \begin{cases}
 0,&p=0,\\[2pt]
 \displaystyle\frac{p(1-\pH)}{(1-p)\pH},&0<p<\pH,\\[6pt]
 1,&\pH\leq p\leq1.
 \end{cases}
 \label{eq:response}
\end{equation}
The response specifies what the agency would do if the administrator
chose $A$ at any belief, including below $\pL$ and above $\pH$,
where the prescribed policy does not use the program.

For every $p\in(0,\pH)$, the response gives
$q=p+(1-p)m(p)=p/\pH$. Verified implementation therefore moves the
posterior to $\pH$, while verified diversion moves it to zero. At
$p\geq\pH$, both types implement under $A$, so a clean inspection
leaves the posterior unchanged. Below $\pH$, the mixture in \eqref{eq:response} makes a clean finding
sufficiently informative to reach the trust threshold. The value of
funding at that threshold makes the strategic agency indifferent
between implementation and diversion, as we show below.

Let $V(p)$ be the administrator's discounted value and $W(p)$ the
strategic agency's discounted value before the current policy choice.
Define the affine expressions
\begin{equation}
 V_A(p)=\frac{(p/\pH)\AH-k-c}{\kappa},
 \qquad V_N(p)=\frac{pb-k}{\delta}.
 \label{eq:branches}
\end{equation}
The associated values are
\begin{equation}
 (V(p),W(p))=
 \begin{cases}
 (0,0),&0\leq p\leq\pL,\\[2pt]
 (V_A(p),R/\kappa),&\pL<p<\pH,\\[2pt]
 (\vH,R/(\beta\tau)),&p=\pH,\\[2pt]
 (V_N(p),R/\delta),&\pH<p\leq1.
 \end{cases}
 \label{eq:values}
\end{equation}

\begin{theorem}\label{thm:eq}
Under \eqref{eq:domain}, the strategies
\eqref{eq:policy}--\eqref{eq:response}, the posterior rules above, and
the values \eqref{eq:values} form a stationary perfect Bayesian
equilibrium. Audited funding and strategic implementation occur with
positive probability at every belief in $(\pL,\pH)$.
\end{theorem}

Appendix~\ref{app:eq} verifies optimality after every current choice
and rules out history-dependent deviations. The theorem establishes
existence of the stated equilibrium. It selects no funding at the lower
cutoff; other choices at that exact belief are also possible, as
discussed below. Figure~\ref{fig:policy} summarizes the funding regions
and the belief transitions.

\begin{figure}[t]
\centering
\begin{tikzpicture}[x=1cm,y=1cm,>=Stealth,font=\small]
\node[anchor=west,font=\small\bfseries] at (0,2.6)
 {(a) Funding at the current belief};
\draw[very thick] (0,1.5)--(12,1.5);
\foreach \x in {0,3.1,8.1,12} \draw (\x,1.35)--(\x,1.65);
\node at (0,1.05) {$0$};
\node at (3.1,1.05) {$\pL$};
\node at (8.1,1.05) {$\pH$};
\node at (12,1.05) {$1$};
\node[align=center] at (1.55,2) {No funding};
\node[align=center] at (5.6,2) {Verification $A$};
\node[align=center] at (10.05,2) {No verification $N$};
\fill (8.1,1.5) circle (2pt);
\node[align=center,font=\footnotesize] at (8.1,0.3)
 {At $\pH$: $A$ with probability $1-\theta$,\\
 $N$ with probability $\theta$};
\node[anchor=west,font=\small\bfseries] at (0,-0.8)
 {(b) Next-period belief from $p\in(\pL,\pH)$};
\node[draw,circle,minimum size=9mm] (p) at (6,-2.4) {$p$};
\node[draw,circle,minimum size=9mm] (zero) at (1.2,-3.5) {$0$};
\node[draw,circle,minimum size=9mm] (high) at (10.8,-3.5) {$\pH$};
\draw[->,thick] (p) to[loop above,looseness=7]
 node[above,font=\footnotesize] {$1-\tau$} (p);
\draw[->,thick] (p) -- node[above,sloped,font=\footnotesize]
 {$\tau(1-p/\pH)$} (zero);
\draw[->,thick] (p) -- node[above,sloped,font=\footnotesize]
 {$\tau p/\pH$} (high);
\node[align=center,font=\footnotesize] at (1.2,-4.3)
 {Verified diversion\\Funding stops};
\node[align=center,font=\footnotesize] at (10.8,-4.3)
 {Verified implementation\\Funding continues};
\end{tikzpicture}
\caption{Equilibrium funding and learning. Cutoff locations are
schematic. At $\pL$ the baseline selects no funding; at $\pH$ the
administrator uses the stated lottery. In panel (b), no inspection
leaves the belief at $p$. Both destination beliefs, $0$ and $\pH$,
are absorbing under equilibrium play.}
\label{fig:policy}
\end{figure}

\subsection{The incentive to implement}

To see why the strategic agency implements, first consider an
unaudited advance. Diversion yields $R$ more today and changes no
public observation, so the agency strictly prefers it under $N$.
Under $A$, diversion can instead be discovered and lead to the loss
of future funding. The payoff difference between implementation and
diversion is
\begin{equation}
 -R+\beta\tau\{W(T_D(p;\mu))-W(0)\}.
 \label{eq:ic}
\end{equation}
Implementation costs the agency the current diversion receipt. It is
worthwhile when this cost is covered by the expected continuation gain
from a clean inspection.

Below $\pH$, the prescribed response makes $T_D=\pH$, and
$\beta\tau W(\pH)=R$. The agency is indifferent and can use the
mixture in \eqref{eq:response}. At $\pH$, certain implementation under
$A$ is also optimal. The administrator's lottery supplies the
continuation rent that makes these choices possible:
\begin{equation}
 W(\pH)=\theta R+\beta W(\pH)
       =\frac{\theta R}{\delta}
       =\frac{R}{\beta\tau}.
 \label{eq:boundaryrent}
\end{equation}
The strategic agency earns a current private payoff only when it
receives an unaudited advance. The frequency $\theta$ of these advances
provides just enough future diversion opportunities to make it willing
to implement under $A$. No promised transfer is needed.

The administrator is also willing to randomize. At $\pH$, its current
expected gain from choosing $A$ instead of $N$ is
$b(1-\pH)-c=0$: the benefit of inducing the strategic type to
implement exactly covers the program cost. Above this belief, the
strategic type is sufficiently unlikely that the additional
implementation no longer justifies verification. The administrator
therefore funds without the program, accepting diversion if the agency
is strategic.

The administrator's value is continuous at both cutoffs:
$V_A(\pL)=0$ and $V_A(\pH)=V_N(\pH)=\vH$. The strategic agency's value is discontinuous, with
\[
 \frac{R}{\kappa}<\frac{R}{\beta\tau}<\frac{R}{\delta}.
\]
This discontinuity matters for incentives. At $\pH$, intermittent
auditing provides the continuation rent in \eqref{eq:boundaryrent}.
Using either one-sided limit instead would destroy the indifference
that supports implementation at lower beliefs. Although $\pH$ is a
single belief, it is reached with positive probability: every clean
inspection from the audited region leads there.

At $\pL$, the administrator is indifferent between stopping and
auditing. It can instead audit with any probability $\pi\in[0,1]$,
leaving the other strategies unchanged and setting
\begin{equation}
 W_\pi(\pL)=\frac{\pi R}{\delta+\beta\pi\tau}.
 \label{eq:lowerselection}
\end{equation}
We use $\pi=0$ in the baseline. This choice matters only when the
initial belief is exactly $\pL$, since a path starting at any other
belief never reaches that cutoff. Appendix~\ref{app:selection}
describes the outcomes when the administrator instead initiates funding
there.

\subsection{The value of access to verification}

Suppose the administrator has only choices $0$ and $N$ in every period.
The strategic agency then always diverts, the posterior remains
unchanged, and the administrator's value is
\begin{equation}
 \widehat V(p)=\frac{\max\{pb-k,0\}}{\delta}.
 \label{eq:benchmark}
\end{equation}
Funding is worthwhile precisely above $k/b$, with indifference at that
belief. This benchmark uses the same infinite horizon and lack of future
commitment as the model with verification.

\begin{proposition}\label{prop:availability}
The equilibrium in Theorem~\ref{thm:eq} gives
$V(p)\geq\widehat V(p)$ at every prior, with strict inequality exactly
on $(\pL,\pH)$.
\end{proposition}

\begin{proof}
The values coincide below $\pL$ and at or above $\pH$. On
$(\pL,k/b]$, $V_A$ is positive and $\widehat V=0$. On
$(k/b,\pH)$,
\begin{equation}
 \delta V_A(p)-(pb-k)
 =\frac{\beta\tau k-\delta c}{\kappa}
       \left(1-\frac{p}{\pH}\right)>0.
 \label{eq:margin}
\end{equation}
The coefficient is positive by \eqref{eq:domain}.
\end{proof}

Access to verification allows the administrator to fund some agencies
whose reputations would be too low for unaudited funding. It also
induces implementation by strategic agencies seeking future advances.
These gains are measured against the permanent absence of verification.
The comparison concerns this particular additional program; a richer
menu can change the equilibrium incentives, as discussed in
Section~\ref{sec:discussion}.

\section{Learning and long-run implementation}\label{sec:longrun}

\subsection{The first inspection and the limiting belief}

Fix an initial prior $x\in(\pL,\pH)$. Until the first inspection, the
administrator chooses $A$ every period, the belief remains $x$, and
implementation has probability $x/\pH$. Let
\begin{equation}
 T=\inf\{t\geq1:p_t\in\{0,\pH\}\}.
 \label{eq:T}
\end{equation}
An inspection in period zero gives $T=1$: the updated belief governs
the next period. Probabilities and expectations with subscript $x$
condition on $p_0=x$ and average over type, unless a type is specified.

Let $f_t$ indicate funding and $d_t$ implementation, with $d_t=0$ when
there is no funding. Realized funding and implemented spending are
\begin{equation}
 r_t=Rf_t,\qquad g_t=Rf_td_t.
 \label{eq:flows}
\end{equation}
Funding is observed, whereas implementation is learned only through
verification. We distinguish the amount advanced, $r_t$, from the
amount used to implement the project, $g_t$. Write $\overline g$
for the limit, when it exists, of $n^{-1}\sum_{t=0}^{n-1}g_t$.

\begin{theorem}\label{thm:longrun}
Starting from $x\in(\pL,\pH)$ in the equilibrium of
Theorem~\ref{thm:eq}, the following statements hold.
\begin{enumerate}[label=(\roman*)]
\item The belief reaches zero or $\pH$ in finite time almost surely
and remains there. The time $T$ has geometric parameter $\tau$, and
\begin{equation}
 \PP_x(p_\infty=\pH)=\frac{x}{\pH},\qquad
 \PP_x(p_\infty=0)=1-\frac{x}{\pH},\qquad
 \E_x T=\frac1\tau.
 \label{eq:limitlaw}
\end{equation}
Conditional on $H$, the destination is $\pH$ with probability one;
conditional on $O$, that probability is $m(x)$.
\item Funding becomes constant after $T$, with
$r_t\to R\ind_{\{p_\infty=\pH\}}$. The implementation average exists
almost surely and has the distribution
\begin{equation}
 \begin{array}{c|ccc}
 \overline g&R&(1-\theta)R&0\\ \hline
 \text{Probability}&x&x(1-\pH)/\pH&1-x/\pH.
 \end{array}
 \label{eq:gdistribution}
\end{equation}
\item Conditional on reaching $\pH$, inspections occur infinitely
often, all report implementation, and the posterior stays at $\pH$.
A surviving strategic agency nevertheless diverts with limiting
frequency $\theta$. Its realized sequence $g_t$ does not converge.
\end{enumerate}
\end{theorem}

The proof is in Appendix~\ref{app:longrun}. The first inspection
determines whether the agency continues to receive funding. Since a
committed agency always implements, it cannot be excluded from this
initially audited region. A strategic agency retains funding with
probability $m(x)$, the probability that it implements the inspected
project. Thus exclusion reveals the strategic type, whereas a clean
inspection leaves both types possible:
\begin{equation}
 \PP_x(H\mid p_\infty=\pH)=\frac{x}{x/\pH}=\pH.
 \label{eq:composition}
\end{equation}

The same argument describes learning before the first inspection. Let
$\mathsf D_z$ denote a point mass at $z$. With probability
$(1-\tau)^t$, no inspection has occurred by date $t$ and the
posterior is still $x$; otherwise it has reached one of the two
absorbing beliefs. Hence, for $t\geq0$,
\begin{equation}
 \mathcal L_x(p_t)=(1-\tau)^t\mathsf D_x+
 [1-(1-\tau)^t]
 \left\{\frac{x}{\pH}\mathsf D_{\pH}
       +\left(1-\frac{x}{\pH}\right)\mathsf D_0\right\}.
 \label{eq:finitelaw}
\end{equation}
It follows that $\E_xp_t=x$ and
\begin{equation}
 \E_x[p_t(1-p_t)]
 =x(1-\pH)+(1-\tau)^t x(\pH-x).
 \label{eq:variance}
\end{equation}
This is the expected posterior variance of the type indicator.
Expected uncertainty declines over time. Its positive limit,
$x(1-\pH)$, comes from agencies that retain funding, since the type
of an excluded agency is known.

At other initial beliefs, reputation does not change under the
baseline policy. If $x\leq\pL$, funding never begins and $p_t=x$.
A committed agency with a sufficiently low initial reputation can
therefore remain unfunded without any adverse finding. If $x=\pH$,
the administrator immediately uses the lottery in \eqref{eq:policy}.
If $x>\pH$, it funds without verification forever; type $H$
implements and type $O$ diverts. Inactivity at a low initial belief
thus differs from exclusion after verified diversion, which reveals
type $O$ and sets the posterior to zero.

\subsection{What a clean audit reveals}

At the trust threshold, both types generate the same public record:
\begin{equation}
 \begin{array}{l|c}
 \text{Observation}&\text{Probability under either type}\\ \hline
 N\text{ and no report}&\theta\\
 A\text{ and no inspection}&(1-\theta)(1-\tau)\\
 A\text{ and verified implementation}&(1-\theta)\tau.
 \end{array}
 \label{eq:publicrecord}
\end{equation}
The two types have identical likelihoods for the administrator's
choices and the inspection outcomes. A clean report therefore leaves
reputation unchanged: implementation was equally likely under either
type. Yet the possibility of a finding of diversion remains essential.
Such a finding would end funding, so the strategic agency continues to
implement under $A$. Auditing sustains conduct even though its
equilibrium reports no longer distinguish types.

Conditional on continued funding, implemented spending averages $R$
for type $H$ and $(1-\theta)R$ for type $O$. Weighting these by
the posterior probability of each type gives
\begin{equation}
 \IH=R[1-\theta(1-\pH)].
 \label{eq:IH}
\end{equation}
This average accounts for diversion of unaudited advances. Clean
findings concern projects assigned to $A$ and provide no evidence
that the agency also implements under $N$.

Before the current choice, expected implemented spending at any belief is
\begin{equation}
 I(p)=
 \begin{cases}
 0,&0\leq p\leq\pL,\\
 (p/\pH)R,&\pL<p<\pH,\\
 \IH,&p=\pH,\\
 pR,&\pH<p\leq1.
 \end{cases}
 \label{eq:Ip}
\end{equation}
Higher reputation need not imply more implementation. In particular,
\[
 \lim_{p\uparrow\pH}I(p)=R>\IH>
 \lim_{p\downarrow\pH}I(p)=\pH R.
\]
The reason is the change in monitoring at the threshold. As the
administrator relaxes verification, the strategic agency gains more
opportunities to divert. Expected implementation falls, but the saving
in program costs offsets this loss in the administrator's payoff.
Its value consequently remains continuous.

\section{Audit cost, detection, and selection}\label{sec:policy}

\subsection{Two technological improvements}

We now compare two changes in verification technology, holding
$b,k,R,\beta$ fixed. A lower $c$ at unchanged $\tau$ makes the
same detection probability available more cheaply. A higher $\tau$
at unchanged $c$ improves detection for the same expected cost. In
each case we compare equilibria of games with different available
programs; the administrator's choices within a game remain $0$, $N$,
and $A$. Both parameter values must satisfy \eqref{eq:domain}.

For an initial prior in the audited region, we use four measures of
the long-run outcome: the probability of continued funding $S$,
expected funding $F$, expected implemented spending $G$, and
residual uncertainty $U$. They are given by
\begin{align}
 S(x)&=\PP_x(p_\infty=\pH)=\frac{x}{\pH},&
 F(x)&=RS(x),\notag\\
 G(x)&=\E_x\overline g
      =Rx\left\{\theta+\frac{1-\theta}{\pH}\right\},&
 U(x)&=\E_x[p_\infty(1-p_\infty)]=\frac{xc}{b}.
 \label{eq:longrunobjects}
\end{align}
Population comparisons below refer to repetitions of this relationship
at a common initial prior. The formula for $G$ averages the three
possible long-run implementation rates in \eqref{eq:gdistribution}. Each rate is a
time average; realized spending by a strategic agency that retains
funding continues to fluctuate between zero and $R$.

\begin{theorem}\label{thm:policy}
Within \eqref{eq:domain}, the equilibrium in Theorem~\ref{thm:eq} has
the following comparative statics.
\begin{enumerate}[label=(\roman*)]
\item At every fixed prior, $V$ is weakly decreasing in $c$ and weakly
increasing in $\tau$. The comparisons are strict for priors in the
audited region under both parameter values.
\item A reduction in $c$ lowers $\pL$ and raises $\pH$. For initial
priors in the common audited region, it lowers $S$, $F$, $G$, and $U$.
\item An increase in $\tau$ at unchanged $c$ lowers $\pL$ and leaves
$\pH$ unchanged. For initial priors in the common audited region,
it raises $G$ and leaves $S$, $F$, and $U$ unchanged.
\end{enumerate}
\end{theorem}

Appendix~\ref{app:policy} proves the comparisons of values and
cutoffs, including cases in which a parameter change moves a fixed
prior across a cutoff. The long-run comparisons follow by
differentiating \eqref{eq:longrunobjects}. For cost changes,
\begin{equation}
 S_c=\frac{x}{b\pH^2}>0,\qquad
 G_c=\frac{Rx(1-\theta)}{b\pH^2}>0,\qquad U_c=\frac xb>0.
 \label{eq:longruncost}
\end{equation}
For detection changes at fixed cost,
\begin{equation}
 S_\tau=U_\tau=0,\qquad
 G_\tau=\frac{Rx(1-\pH)\delta}{\pH\beta\tau^2}>0.
 \label{eq:longrundetection}
\end{equation}

\subsection{Why cheaper auditing can reduce implementation}

When verification becomes cheaper, the administrator is willing to
use it at higher reputations. The threshold for relaxing monitoring,
$\pH=1-c/b$, therefore rises. A clean inspection must now provide
stronger evidence of commitment to bring an agency to that threshold.
Since a committed agency always implements, this requires a strategic
agency in the audited region to implement less often:
\begin{equation}
 m(x)=\frac{xc}{(1-x)(b-c)}.
 \label{eq:mc}
\end{equation}
The lower implementation probability makes a clean finding sufficiently
informative to reach the new trust threshold. The continuation rent
there still makes the strategic agency indifferent. Every committed
agency in the initial audited region retains funding, so the fall in
$S$ comes entirely from fewer strategic agencies passing the first
inspection.

Agencies that retain funding are consequently more likely to be
committed: their posterior probability of type $H$ is $\pH$.
Expected implementation per funded period, $\IH$, rises as $c$
falls, even though a strategic agency that reaches the threshold still
implements with the same frequency $1-\theta$. The improvement among
funded agencies comes from their composition. Fewer strategic agencies
receive advances, and their exclusion reduces implementation in the
initial population.

The administrator gains because it pays less for verification and is
less likely to fund indefinitely an agency that diverts unaudited
advances. Its value accounts for the cost of those
advances as well as the benefit of implementation. The measure $G$
records only implemented expenditure. The two criteria can therefore
rank the same monitoring change differently.

The decline in implemented spending applies to agencies whose initial
reputations lie in the audited region under both cost levels. Lower
cost also allows funding to begin at some reputations where it was
previously withheld. For these agencies, or for others whose initial
belief crosses a cutoff, the long-run spending comparison can differ.
The gain in the administrator's value holds at every fixed prior.

Better detection at fixed cost leaves $m(x)$, $\pH$, and the
probability of continued funding unchanged. Its effect is on timing
and incentives: the first inspection arrives sooner, and a lower
frequency $\theta=\delta/(\beta\tau)$ of unaudited advances suffices
to induce implementation at the trust threshold. Strategic agencies
that retain funding then implement more often, while their type
remains uncertain.

Inspection frequency and learning also need to be distinguished. The
expected long-run frequency of inspections is
\begin{equation}
 J(x)=\frac{x}{\pH}(1-\theta)\tau
     =\frac{x}{\pH}\left(\tau-\frac{\delta}{\beta}\right).
 \label{eq:J}
\end{equation}
For common audited priors, cheaper verification lowers $J$ because
fewer agencies retain funding. Better detection at unchanged cost
raises it. In both cases, inspections after the trust threshold has
been reached sustain implementation without further revealing type.
Their frequency alone therefore measures neither remaining uncertainty
nor the value of verification.

\subsection{When inspections must be paid for}

Suppose instead that each inspection costs $\gamma>0$, so the
expected program cost is $c=\gamma\tau$. A higher $\tau$ now
improves detection and raises expected cost at the same time. The
administrator need not gain from this change. For a prior remaining
in the audited region, Appendix~\ref{app:priced} gives
\begin{equation}
 \frac{dV_A(p;\gamma\tau,\tau)}{d\tau}
 =\frac{\beta k-\gamma\delta}{\kappa^2}
       \left(1-\frac{p}{p^\dagger}\right),
 \qquad
 p^\dagger=\frac{\delta(b-\gamma\tau)^2}
                  {b(\delta b+\gamma\beta\tau^2)}<\pH.
 \label{eq:priced}
\end{equation}
The coefficient preceding the parentheses is positive under the
parameter restrictions. Within the audited region, the administrator
therefore gains from greater detection below $p^\dagger$ and loses
above it. Both intervals are nonempty when inspection cost is
sufficiently small and positive. Near the trust threshold, little
information remains to be gained and the additional cost dominates.

For common audited priors, continued funding becomes more likely and
expected long-run implementation rises in this comparison. The higher
program cost lowers $\pH$, allowing more strategic agencies to retain
funding and leaving greater uncertainty about type. More inspections
can thus be associated with both higher implementation and a lower
value to the administrator. The cost of obtaining those inspections
is essential to their assessment.

\section{Discussion and conclusion}\label{sec:discussion}

We have studied a repeated funding relationship in which an agency's
reputation determines its access to advances and the monitoring it
faces. Cheaper auditing makes funding worthwhile at lower initial
reputations, but also raises the reputation required for continued
funding with intermittent verification. For agencies initially audited
under both cost levels, fewer strategic agencies retain funding.
Expected long-run implemented spending falls, even as implementation
per funded agency and the administrator's value rise. For initial
beliefs audited under both programs, better detection at unchanged cost
instead raises implementation without changing the probability that
either type retains funding.

The conclusions depend on how information arrives. An inspection may
be inconclusive without changing the mechanism, provided this outcome
is independent of type and action. The effective probability of a
conclusive finding then replaces the baseline detection probability;
Appendix~\ref{app:verification} gives the corresponding equilibrium.
False accusations have a different effect. If both types implement
and every report is possible after either action, reports leave the
posterior unchanged. Under stationary strategies based on this
posterior, diversion then yields a current gain without a continuation
loss. Certain implementation by the strategic type at the trust
threshold can no longer be sustained. An equilibrium with partial
implementation would require a separate analysis.

The fixed verification menu and the agency's observation of the
current program also matter. Adding an intermediate audit intensity
can make the prescribed continuation values inconsistent with an
optimal agency response. Concealing whether $A$ or $N$ was chosen
prevents the agency from conditioning implementation on that choice.
Appendix~\ref{app:limits} shows why either change requires a new
equilibrium construction. Our results concern the specified funding
instruments and do not characterize an optimal monitoring contract.

Finally, persistent uncertainty depends on the absence of information
outside the verification program. Reports about unaudited projects
would alter both beliefs and incentives. Recurring independent news
about the fixed type would eventually reveal it, even after funding
stops. The absorbing reputations in our model rely on exact verification
of diversion and uninformative unaudited funding. The first feature
violates full support, and the second precludes identification of
actions under $N$. The monitoring assumptions behind the impermanence
results of \citet{CMS2004} therefore do not hold here.

The model suggests that a monitoring reform should be assessed through
its effects on continued funding as well as on the use of each advance.
The agencies that remain funded can deliver more per project while
aggregate implementation in the original population declines. Such a
decline need not indicate that the reform has reduced the value of
delegation to the administrator.

\appendix
\section{Equilibrium verification}\label{app:eq}

\subsection{Ordering of the cutoffs}

The bound $c<b-k$ implies $k/b<\pH<1$ and $\vH>0$. Moreover,
\begin{equation}
 k\AH-b\pH(k+c)
 =\frac{b-k-c}{\delta}\{\beta\tau k-\delta c\}>0.
 \label{eq:cutoffproof}
\end{equation}
Thus $\pL<k/b$. Positivity of all terms defining $\pL$ and
$\beta\tau>\delta$ complete \eqref{eq:ordering}.

\subsection{The agency's response}

Fix the posterior rule generated by the prescribed conjecture. Under
$A$, implementation and diversion yield respectively
\begin{align}
 U_D&=\beta\{(1-\tau)W(p)+\tau W(T_D)\},\notag\\
 U_X&=R+\beta\{(1-\tau)W(p)+\tau W(0)\}.
 \label{eq:agencyreturns}
\end{align}
For a deviating implementation probability $\alpha\in[0,1]$, the
return is $\alpha U_D+(1-\alpha)U_X$, with the conjecture $\mu$
in $T_D$ held fixed. An interior choice is optimal exactly when
$U_D=U_X$; in equilibrium, $\alpha=\mu=m(p)$.
For $0<p<\pH$, \eqref{eq:response} gives $T_D=\pH$ and
$\beta\tau W(T_D)=R$, so the agency is indifferent. This also
verifies its response if the administrator orders an audit below
$\pL$. At $p=\pH$, certain implementation leaves the posterior at
$\pH$ and the same indifference holds. For $p>\pH$,
\[
 \beta\tau W(p)=\frac{\beta\tau R}{\delta}>R,
\]
so implementation after $A$ is strictly optimal. At $p=0$, all reports
lead to zero and diversion is strictly preferred. Under $N$, diversion
strictly dominates implementation at every belief because the
continuation is unchanged.

The agency's value equations are also satisfied. Below and at $\pL$,
no funding gives $W=\beta W=0$. In the audited region, indifference
allows the value to be computed using diversion:
\[
 W(p)=R+\beta(1-\tau)W(p)=R/\kappa.
\]
At $\pH$, \eqref{eq:boundaryrent} gives the required value. Above
$\pH$, unaudited funding gives $W=R+\beta W=R/\delta$.

\subsection{The administrator's choice}

Given the agency's response after every choice, the administrator's
returns are
\begin{align}
 F_0(p)&=\beta V(p),\notag\\
 F_N(p)&=pb-k+\beta V(p),\notag\\
 F_A(p)&=qb-k-c+\beta\{(1-\tau)V(p)
             +\tau qV(T_D)+\tau(1-q)V(0)\}.
 \label{eq:authorityreturns}
\end{align}
For $0<p<\pH$, substitution gives
\begin{equation}
 F_A(p)=\frac{p}{\pH}\AH-k-c+\beta(1-\tau)V(p).
 \label{eq:FA}
\end{equation}
For $p\leq\pL$, the prescribed value is zero. The return from $A$
is nonpositive by the definition of $\pL$, including $-k-c$ at
$p=0$. The return from $N$ is strictly negative because
$\pL<k/b$. Stopping is optimal.

For $\pL<p<\pH$, $V_A>0$ and \eqref{eq:FA} gives $F_A=V_A$.
Stopping yields only $\beta V_A<V_A$. Identity~\eqref{eq:margin}
gives $V_A-F_N>0$. Thus $A$ is strictly optimal in this interval.
At $p=\pH$, $A$ and $N$ both yield current expected payoff
$b-k-c$ and preserve the posterior on the equilibrium path. Both
attain $\vH$, while stopping is strictly worse. Hence the
administrator is willing to use the prescribed lottery.

At $p>\pH$, choosing $N$ gives $V_N>0$. If the administrator
deviates to $A$, both types implement and the on-path posterior remains
$p$. The gain over $N$ is
\[
 F_A(p)-V_N(p)=b(1-p)-c<0.
\]
Stopping is strictly worse as well. This verifies the choice at every
belief, including the boundary $p=1$.

\subsection{Longer deviations}

The preceding inequalities verify optimal current actions at every
decision history represented by the public posterior, including after
choices that are not taken on the equilibrium path. Current mixing
cannot improve on the best pure deviation because returns are linear
in deviation probabilities.

The values satisfy $0\leq V(p)\leq(b-k)/\delta$ and
$0\leq W(p)\leq R/\delta$ at every belief. For the administrator,
iterate $V(p)\geq F_u(p)$ along any alternative
sequence of choices, using the prescribed agency response after each
realized choice. Its expected payoff over the first $n$ periods plus
$\beta^nV(p_n)$ is bounded above by its initial value. Boundedness of
$V$ makes the terminal term vanish as $n\to\infty$. The same
argument applies to the strategic agency, first using its action
inequality after the observed policy choice and then averaging over
that choice. Its terminal term $\beta^nW(p_n)$ also vanishes.
These arguments allow history-dependent deviations and private
randomization. The prescribed strategies attain the value equations;
together with the posterior rules, this proves Theorem~\ref{thm:eq}.

\section{Learning and spending}\label{app:longrun}

\subsection{Proof of Theorem~\ref{thm:longrun}}

Before the first inspection, the belief stays at $x$ and $A$ is chosen
each period. For any integer $n\geq1$,
\begin{align}
 \PP_x(T=n,p_T=\pH)
  &=(1-\tau)^{n-1}\tau\frac{x}{\pH},\notag\\
 \PP_x(T=n,p_T=0)
  &=(1-\tau)^{n-1}\tau\left(1-\frac{x}{\pH}\right).
 \label{eq:jointtime}
\end{align}
The inspection draw is independent of type and of the current action.
Conditional on inspection, $H$ implements with probability one and $O$
with probability $m(x)$; averaging gives $x/\pH$. The geometric tail
tends to zero and its mean is $1/\tau$. The destination is therefore
reached almost surely. Summing \eqref{eq:jointtime} proves the
unconditional terminal law, and conditioning on type gives the stated
survival probabilities. The same calculation establishes independence
of inspection time and terminal destination.

After verified diversion, funding stops at posterior zero. After
verified implementation, the posterior is $\pH$. Under $A$, both
types implement there, while $N$ generates no informative report.
The likelihoods in \eqref{eq:publicrecord} are identical for every
finite future public history, so the posterior remains $\pH$.
These observations also prove \eqref{eq:finitelaw}; taking its first
two moments gives \eqref{eq:variance}.

At $\pH$, fresh independent draws imply frequency $\theta$ for
$N$, frequency $1-\theta$ for $A$, and frequency
$(1-\theta)\tau$ for actual inspections, almost surely. The last
frequency is strictly positive, so inspections occur infinitely often
and all record implementation. Type $H$ implements every project;
type $O$ implements precisely under $A$. The finite history before
$T$ has zero weight in time averages. Hence $\overline g=R$ for
type $H$, $\overline g=(1-\theta)R$ for a surviving strategic agency,
and $\overline g=0$ for an excluded one. Their probabilities are
\[
 x,\qquad (1-x)m(x)=\frac{x(1-\pH)}{\pH},\qquad
 1-\frac{x}{\pH},
\]
which proves \eqref{eq:gdistribution}. Both funding arrangements occur
infinitely often at $\pH$. A surviving strategic agency's $g_t$ consequently
takes both zero and $R$ infinitely often and has no period-by-period
limit. Funding itself is constant after $T$.

\subsection{Selection at the lower cutoff}\label{app:selection}

Choose $A$ with probability $\pi\in[0,1]$ at $\pL$ and $0$
otherwise, leaving all other strategies unchanged. The administrator's
value is still zero. The strategic value solves
\[
 W_\pi(\pL)=\pi R+
       \beta(1-\pi\tau)W_\pi(\pL),
\]
which gives \eqref{eq:lowerselection}. Since this value is at most
$R/\kappa<R/(\beta\tau)$, the response comparisons elsewhere are
unchanged. At the cutoff, the audit response still gives
$T_D=\pH$ and satisfies the same indifference condition.

For $\pi=0$, funding never starts. For any $\pi>0$, the exit time
has geometric parameter $\pi\tau$ and
\begin{equation}
 \PP_{\pL}(p_\infty=\pH)=\frac{\pL}{\pH},\qquad
 \PP_{\pL}(p_\infty=0)=1-\frac{\pL}{\pH},\qquad
 \E_{\pL}T=\frac1{\pi\tau}.
 \label{eq:selectionlaw}
\end{equation}
The proof repeats \eqref{eq:jointtime} with inspection probability
$\pi\tau$. Every positive audit probability gives the same limiting distribution
but a different expected delay. As $\pi$ tends to zero, this delay
diverges; at $\pi=0$, the belief never changes. Since no other
initial belief leads to $\pL$, the choice among these equilibria
affects only the path starting at that exact prior.

\section{Policy comparisons}\label{app:policy}

\subsection{Value and cutoff derivatives}

On the audited branch, differentiation gives
\begin{align}
 (V_A)_c(p)
 &=\frac1\kappa\left\{
       \frac{pb[b-\beta\tau k/\delta]}{(b-c)^2}-1\right\},
 \label{eq:Vc}\\
 (V_A)_\tau(p)
 &=\frac{\beta(1-p/\pH)(k+c)}{\kappa^2}.
 \label{eq:Vtau}
\end{align}
For \eqref{eq:Vc}, use the identity
\[
 \frac{\AH}{\pH}
 =b\left\{\frac{\beta\tau}{\delta}
       +\frac{b-\beta\tau k/\delta}{b-c}\right\}.
\]
For \eqref{eq:Vtau}, put $q=p/\pH$. At fixed $c$, $q$ is
constant, $(\AH)_\tau=\beta\vH$, and $\kappa_\tau=\beta$, so
\begin{align*}
 \kappa^2(V_A)_\tau
 &=\beta\{q\vH\kappa-q\AH+k+c\}\\
 &=\beta\{q(\delta\vH-b)+k+c\}
 =\beta(1-q)(k+c).
\end{align*}

The detection derivative is positive for $p<\pH$. To sign the cost
derivative, put $D=b-\beta\tau k/\delta$. If $D\leq0$, the
sign is immediate. If $D>0$, the cost restriction gives $D<b-c$.
Since $p<\pH=(b-c)/b$, we obtain
$pbD/(b-c)^2<D/(b-c)<1$. Thus $(V_A)_c<0$ throughout the audited
region and at its lower boundary.

The upper cutoff has derivatives $(\pH)_c=-1/b$ and
$(\pH)_\tau=0$. The lower cutoff satisfies
\begin{equation}
 (\pL)_\tau=-\frac{\beta\vH\pH(k+c)}{\AH^2}<0,
 \qquad
 (\pL)_c=-\frac{\pH\kappa}{\AH}(V_A)_c(\pL)>0,
 \label{eq:cutoffderivatives}
\end{equation}
where the second expression follows by differentiating
$V_A(\pL)=0$.

Outside the audited region, the administrator's value is either zero
or $(pb-k)/\delta$, both independent of $c$ and $\tau$. These
branches meet $V_A$ continuously. For a fixed prior, following a
parameter change through the moving cutoffs therefore preserves the
weak derivative ordering. On a common audited interval it is strict.
This proves parts (i) and the cutoff claims of Theorem~\ref{thm:policy}.

\subsection{Long-run derivatives}

At fixed $\tau$, $\theta$ is unchanged and $(\pH)_c=-1/b$.
Differentiating \eqref{eq:longrunobjects} gives
\eqref{eq:longruncost}, and $F_c=RS_c>0$. At fixed $c$,
$(\pH)_\tau=0$ and
$\theta_\tau=-\delta/(\beta\tau^2)$. This gives
\eqref{eq:longrundetection} and $F_\tau=0$, completing the theorem.
For inspections, \eqref{eq:J} gives $J_c>0$ and
$J_\tau=x/\pH>0$. These formulas apply while the initial prior
remains strictly within the audited region under both parameter
values.

\subsection{Priced inspections}\label{app:priced}

With $c=\gamma\tau$, the parameter restrictions become
\begin{equation}
 \tau>\delta/\beta,\qquad \gamma\tau<b-k,\qquad
 0<\gamma<\beta k/\delta,\qquad \tau\leq1.
 \label{eq:pricedomain}
\end{equation}
Let $C=\beta k-\gamma\delta>0$ and $y=b-\gamma\tau>0$.
The numerator of the chain-rule derivative,
\[
 D(p)=\kappa^2\{(V_A)_\tau+\gamma(V_A)_c\},
\]
is affine in $p$. Direct substitution into
\eqref{eq:Vc}--\eqref{eq:Vtau} gives
$D(0)=C$ and
$D(\pH)=-\gamma\kappa\tau C/(\delta y)$. Consequently,
\begin{align*}
 D(p)&=C\left\{1-\frac{p}{\pH}
                   \left(1+\frac{\gamma\kappa\tau}{\delta y}\right)
             \right\}\\
     &=C(1-p/p^\dagger),
\end{align*}
because $\delta y+\gamma\kappa\tau=\delta b+\gamma\beta\tau^2$.
This proves \eqref{eq:priced}. Moreover,
\[
 \frac{p^\dagger}{\pH}
 =\frac{\delta(b-\gamma\tau)}{\delta b+\gamma\beta\tau^2}<1.
\]
For fixed $b,k,\beta,\tau$ satisfying the other restrictions, as
$\gamma\downarrow0$, both $p^\dagger$ and $\pH$ tend to one,
whereas $\pL$ tends to
$k/[b+\beta\tau(b-k)/\delta]<1$. Hence
$\pL<p^\dagger<\pH$ for sufficiently small positive $\gamma$.
If instead $p^\dagger\leq\pL$, the value derivative is negative
throughout the audited region.

The long-run formulas become
\begin{equation}
 S(x)=\frac{x}{1-\gamma\tau/b},\qquad
 G(x)=\frac{Rx[1-\delta\gamma/(\beta b)]}{1-\gamma\tau/b},
 \qquad U(x)=\frac{x\gamma\tau}{b}.
 \label{eq:pricedlongrun}
\end{equation}
The numerator of $G$ is positive since
$\delta\gamma/\beta<k<b$. All three expressions increase with
$\tau$ for common audited priors.

\section{Verification and information}\label{app:verification}

\subsection{Inconclusive inspections}

Suppose an attempted inspection is conclusive with probability
$\eta\in(0,1]$, independently of type and action. A conclusive
finding reveals the action exactly. Otherwise the finding conveys no
information. Retain $c$ as the total expected cost, including
unsuccessful attempts, and put $\nu=\tau\eta$.

\begin{proposition}\label{prop:erasure}
If $\beta\nu>\delta$ and
$0<c<\min\{b-k,\beta\nu k/\delta\}$, the equilibrium and learning
results hold with $\tau$ replaced by $\nu$. At the trust threshold,
the frequency of inspection attempts is $(1-\theta_\nu)\tau$ and
the frequency of conclusive findings is $(1-\theta_\nu)\nu$, where
$\theta_\nu=\delta/(\beta\nu)$.
\end{proposition}

\begin{proof}
No attempt and an inconclusive attempt both leave the posterior
unchanged. Their combined probability is $1-\nu$, independently of
type and action. The probabilities of verified implementation and
diversion are exactly those generated by the baseline with
verification probability $\nu$. The incentive difference becomes
$-R+\beta\nu[W(T_D)-W(0)]$, while the administrator still pays $c$.
Every value equation and deviation comparison therefore has the
original form with $\nu$ in place of $\tau$. The first informative
event is now the first conclusive inspection. Multiplying the program
frequency by the probability of an attempt or a conclusive finding
gives the stated frequencies.
\end{proof}

\subsection{Errors, menus, and disclosure}\label{app:limits}

We consider three changes under which the equilibrium constructed
above cannot be retained. The arguments concern its particular
strategies and values; other equilibria in the modified games may
exist.

First, suppose an inspection returns a positive report with probability
$1-\varepsilon$ after implementation and $\varepsilon$ after
diversion, where $0<\varepsilon<1/2$. If both types implement with
probability one at an interior belief, both reports have the same
distribution under the two types. Bayes' rule leaves the posterior at
$p$ after either report and after no inspection. Under stationary
posterior strategies, a deviation to diversion changes report
probabilities but not continuation value and earns an additional $R$.
Certain implementation by the strategic type is therefore impossible
after an audited choice at such a belief. This rules out the exact
pooling behavior in Theorem~\ref{thm:eq}, not partial-implementation
equilibria or limits of equilibria as $\varepsilon\to0$.

Second, add an announced program with verification probability
$a\in(\delta/\beta,\tau)$, keeping the baseline continuation
values. At $p=\pH$, any conjecture $\mu<1$ implies
\[
 T_D=\frac{\pH}{\pH+(1-\pH)\mu}>\pH.
\]
The incentive difference is then
$R(\beta a/\delta-1)>0$, so the agency chooses certain
implementation, contradicting $\mu<1$. But $\mu=1$ gives
$T_D=\pH$ and incentive difference $R(a/\tau-1)<0$.
No response to the new choice is consistent with these continuation
values, regardless of its cost. Even an unused feasible choice must
have an optimal response. A new equilibrium would therefore require
more than adding that choice to the old policy.

Third, conceal the realized choice between $A$ and $N$ until after
the agency acts. Its baseline contingent actions at $\pH$ are then
infeasible. Under the old lottery and continuation value, certain
implementation would have incentive difference
\[
 -R+\beta(1-\theta)\tau W(\pH)=-\theta R<0.
\]
Thus the old allocation cannot be preserved by simply hiding the
current program choice. This is distinct from the hidden inspection
draw under an announced program, which is already part of the baseline.

\subsection{Other information}

An informative report under $N$ would also change the baseline
transition law. If, under the old actions, a positive report occurs
with probability $1-\varepsilon$ after implementation and
$\varepsilon$ after diversion, with $0<\varepsilon<1/2$, then
\[
 T_+(p)=\frac{p(1-\varepsilon)}
 {p(1-\varepsilon)+(1-p)\varepsilon}>p,
 \qquad
 T_-(p)=\frac{p\varepsilon}
 {p\varepsilon+(1-p)(1-\varepsilon)}<p.
\]
The trust threshold is no longer absorbing under those actions.
The strategic agency's incentive under $N$ also changes, so this
calculation alone does not establish complete learning in an equilibrium
of the modified game.

A sharper conclusion holds for news about the fixed type that is
independent of behavior. Suppose news arrives independently in each
period with probability $\xi>0$, even without funding. Conditional on
an arrival, the probability of a positive signal is $\rho_H$ under
$H$ and $\rho_O$ under $O$, with $0<\rho_O<\rho_H<1$. The signals
are independent over time conditional on type and independent of
actions. There are infinitely many arrivals almost surely, and the
frequency of positive news converges to different limits under the two
types. The type is therefore measurable with respect to the infinite
public history. If $\mathcal F_t$ denotes the full public information
at time $t$, convergence of conditional expectations gives
\[
 p_t=\E[\ind_{\{H\}}\mid\mathcal F_t]
 \longrightarrow\ind_{\{H\}}\quad\text{almost surely}.
\]
This learning conclusion holds for Bayesian play from an interior
prior without specifying equilibrium behavior in the augmented game.
Recurring independent information eventually identifies type, whereas
the baseline reports depend on actions that verification itself can
induce.

\end{document}